\documentclass[prd,floats,floatfix,nofootinbib]{revtex4}
\usepackage[dvips]{graphicx}
\usepackage{graphics}
\usepackage{dcolumn}
\usepackage{amssymb}
\usepackage{bm}
\usepackage{amsmath}
\usepackage{color}
\usepackage{epstopdf}

\def\spose#1{\hbox to 0pt{#1\hss}}

\def\lta{\mathrel{\spose{\lower 3pt\hbox{$\mathchar"218$}}
     \raise 2.0pt\hbox{$\mathchar"13C$}}}
\def\gta{\mathrel{\spose{\lower 3pt\hbox{$\mathchar"218$}}
     \raise 2.0pt\hbox{$\mathchar"13E$}}}
\newcommand{\be}{\begin{equation}}
\newcommand{\en}{\end{equation}}
\newcommand{\bea}{\begin{eqnarray}}
\newcommand{\ena}{\end{eqnarray}}

\begin{document}

\begin{center}
{\bf
Essay written for the Gravity Research Foundation 2020 Awards for Essays on Gravitation}
\end{center}

\title{Nonsingular Black Holes From Charged Dust Collapse:\\
A Concrete Mechanism to Evade Interior Singularities in General Relativity}

\author{Rodrigo Maier\footnote{rodrigo.maier@uerj.br}
\vspace{0.5cm}}

\affiliation{Departamento de F\'isica Te\'orica, Instituto de F\'isica, Universidade do Estado do Rio de Janeiro,\\
Rua S\~ao Francisco Xavier 524, Maracan\~a,\\
CEP20550-900, Rio de Janeiro, Brazil\\
}


\date{\today}

\begin{abstract}
In this essay we examine the gravitational collapse of a nonrelativistic charged perfect fluid
interacting with a dark energy component. Given a simple factor for the energy transfer, 
we obtain a nonsingular interior solution which naturally matches the Reissner-Nordstr\"om-de Sitter exterior geometry.
We also show that the interacting parameter is proportional to the overall charge of the final black hole thus formed. 
For the case of quasi-extremal configurations, we propose a statistical
model for the entropy of the collapsed matter. This entropy extends
Bekenstein's geometrical entropy by an additive constant proportional to the area of the extremal
black hole.
\end{abstract}
\maketitle
\newpage
Although General Relativity is the most successful theory that currently describes gravitation, it presents
intrinsic pathologies when one tries to construct a concrete gravitational collapse model which may be
understood as an interior solution of a black hole. Apart from instabilities in the exterior spacetime\cite{matzner,maeda}, the existence of singularities within the event horizon is a puzzle which permeates all the four classical black holes solutions in general relativity\cite{chandra}. Neglecting charge and angular-momentum, the Schwarzschild black hole
is the only configuration which allows one to obtain an interior -- though singular -- solution\cite{oppen}. All others lack of proper interior solutions and interior singularities are manifested. 

The issue of a cosmological constant in black hole physics has been 
a subject of interest over the last years\cite{mo}-\cite{perlick}. 
However, from the cosmological point of view
it has been shown that an interacting dark energy component may relieve some cosmological tensions of observational 
data\cite{salvatelli}-\cite{kumar}. In this sense, a question which naturally arises is what would be
the consequences of assuming an interacting dark energy in gravitational collapse processes which may engender black holes. 
This essay is devoted to examine this question.

We start by considering the action
\begin{eqnarray}
\label{eq1}
\nonumber
S=\frac{1}{2{\kappa}^2}\int\sqrt{-g}(R+2\Lambda_I)d^4x+\int\sqrt{-g}({\cal L}_I+{\cal L}_\gamma) d^4x,
\end{eqnarray}
where $\kappa^2$ is the Einstein constant, $g$ is the determinant of a $4$-dimensional geometry and
$R$ is the Ricci scalar. 
We denote $\Lambda_I$ as a dark energy component
which interacts with a nonrelativistic perfect fluid whose lagrangian is given by ${\cal L}_I$. 
Finally, ${\cal L}_\gamma$ stands for radiation.

Variations of this action with respect to the metric furnish the field equations subjected to the following Bianchi 
identities 
\begin{eqnarray}
\label{eq5m1}
\kappa^2 \nabla_\mu {^{(I)}}T^\mu_{~~\nu}&=&Q_\nu,\\
\label{eq51}
\nabla_\nu \Lambda_I &=& -Q_\nu,
\end{eqnarray}
where ${^{(I)}}T^\mu_{~~\nu}$ is the energy-momentum tensor which comes from ${\cal L}_I$. $Q_\nu$ on the other hand, represents the energy-momentum transfer between dark energy and the nonrelativistic perfect fluid.

Let us now consider a FLRW geometry in comoving coordinates $(r, \theta,\varphi)$ given by
\begin{eqnarray}
\label{eq6}
ds^2=dt^2-a^2(t)\Big[\frac{dr^2}{1-kr^2}+r^2(d\theta^2+\sin^2{\theta}d\varphi^2)\Big].
\end{eqnarray}
%
In the following we shall restrict ourselves to the geodesic scenario\cite{wands}
so that $Q^\mu =Q V^\mu$, with $V^\mu\equiv \delta^\mu_{~t}$.
It can then be shown that equations (\ref{eq5m1}) and (\ref{eq51})
furnish
\begin{eqnarray}
\label{eq11}
\kappa^2[\dot{\rho}_I+3H\rho_I]&=&Q,\\
\label{eq111}
\dot{\Lambda}_I&=&- Q,
\end{eqnarray}
In order to solve the dynamics we further assume the simple form for the energy transfer
\begin{eqnarray}
\label{eq14}
Q=\xi   (\Lambda_I-\Lambda_0) H,
\end{eqnarray}
where $\Lambda_0$ is the usual cosmological constant and $H\equiv \dot{a}/{a}$.
In this case, from (\ref{eq11}) and (\ref{eq111}) we obtain
\begin{eqnarray}
\Lambda_I=\Lambda_0+\kappa^2\frac{\lambda}{a^\xi},~~\rho_I=\frac{E_d}{a^3}-\frac{\xi\lambda}{(\xi-3)a^\xi},
\end{eqnarray}
where $\lambda$ and $E_d$ are positive constants of integration. 
Furthermore, it is easy to see that Einstein field equations have a first integral given by
\begin{eqnarray}
\label{eqn2}
\frac{\dot{a}^2}{2}+V(a)=0,
\end{eqnarray}
where
\begin{eqnarray}
\label{eqn3}
V(a)\equiv \frac{k}{2}-\frac{1}{6}\Big[\Lambda_0 a^2 + \kappa^2 \Big(\frac{E_d}{a} +\frac{E_\gamma}{a^2} \Big)\Big]+\frac{\kappa^2\lambda a^{2-\xi}}{2(\xi-3)},
\end{eqnarray}
and $E_\gamma$ is a positive constant of integration connected to the radiation energy density.
In the above we see that as long as $\xi\geq 4$ an infinite potential barrier may avoid the classical singularity predicted by General Relativity. 
In order to fix the parameter $\xi$, we consider the matching of the interior geometry (\ref{eq6}) with
the exterior spacetime at the surface $r \equiv \gamma={\rm constant}$. 
To this end let us assume that $(\bar{t}, \bar{r}, \bar{\theta}, \bar{\phi})$
are new coordinates defined by 
\begin{eqnarray}
\label{trans1}
\bar{t}:=\chi(\psi(t,\bar{r})),~~\bar{r}=ar,~~\bar{\theta}=\theta,~~\bar{\phi}=\phi,
\end{eqnarray}
where the function $\psi$ satisfies
\begin{eqnarray}
\label{gtcn}
\Big(\frac{\partial \psi}{\partial t}\Big)^{-1}\Big(\frac{\partial \psi}{\partial \bar{r}}\Big)=\frac{a\dot{a}\bar{r}}{a^2-\bar{r}^2(k+\dot{a}^2)}.
\end{eqnarray}
Therefore, we ensure that the metric is diagonal and
\begin{eqnarray}
\label{grr}
g_{\bar{r}\bar{r}}=-\Big[1-\frac{\bar{r}^2}{a^2}(k+\dot{a}^2)\Big]^{-1}\Big|_{r=\gamma}.
\end{eqnarray}
It can be easily shown that a solution for the partial differential equation (\ref{gtcn}) is given by
\begin{eqnarray}
\label{matching1}
\psi(t,\bar{r})=\delta+\frac{\mu}{a}\sqrt{a^2-k\bar{r}^2}\exp{\Big(-k\int\frac{1}{a\dot{a}^2}\Big)},
\end{eqnarray}
where $\delta$ and $\mu$ are arbitrary constants.
Furthermore, employing the integrating factor technique, we define the function $\chi[\psi(t, \bar{r})]$
by the differential equation
\begin{eqnarray}
\label{matching2}
\frac{d\chi}{d\psi}=\frac{a^2\dot{a}}{a^2-\bar{r}(k+\dot{a}^2)}\big[\frac{\sqrt{a^2-k\bar{r}^2}}{k(\psi-\delta)}\Big]\Big|_{r=\gamma},
\end{eqnarray}
where $a(t)$ is an implicit function of $\psi$. In fact, for the physical domain of parameters to be
considered, it can be shown that $\psi$ is a monotonous function of $a$
which can be properly inverted. In this case we obtain
\begin{eqnarray}
g_{\bar{t}\bar{t}}=-\frac{1}{g_{\bar{r}\bar{r}}}\Big|_{r=\gamma}.
\end{eqnarray}
Finally, fixing the initial conditions
\begin{eqnarray}
\label{ic}
a(0)=1,~~\dot{a}(0)=0,
\end{eqnarray}
it can be shown that
\begin{eqnarray}
\label{extg}
ds^2=F(\bar{r})d\bar{t}^2-\frac{1}{F(\bar{r})}d\bar{r}^2-\bar{r}^2(d\bar{\theta}^2+\sin^2{\bar{\theta}}d\phi^2),
\end{eqnarray}
where
\begin{eqnarray}
\label{grr}
F(\bar{r})\equiv 1-\frac{2GM}{\bar{r}}+\kappa^2\Big[\frac{\lambda \gamma^\xi}{(\xi-3)\bar{r}^{\xi-2}}-\frac{E_\gamma\gamma^4}{3\bar{r}^2}\Big]-\frac{\Lambda_0}{3}\bar{r}^2,
\end{eqnarray}
and $M\equiv 4\pi\gamma^3 E_d/3$. At this stage, it is important to draw the reader's attention to a word to note. Let us consider an observer external to a spherically symmetric cloud of charged dust which collapses due to the action of its own gravity. As the collapse starts, charged particles are accelerated and a flow of electromagnetic radiation is ejected to the exterior spacetime towards our observer. This process continues until an event horizon is formed and such flow ceases completely. Therefore, after a finite amount of time
our external observer can not detect anything but the electric field due to the charge of matter distribution. In other words, after a sufficient amount of time, our external observer has to be embedded in an Reissner-Nordstr\"om-de Sitter spacetime. 
From (\ref{grr}) we see that the exterior solution restores the Reissner-Nordstr\"om-de Sitter metric as long as $\xi=4$. In this case,
\begin{eqnarray}
\label{rn}
F(\bar{r})=1-\frac{2GM}{\bar{r}}+\frac{\beta^2_q}{\bar{r}^2}-\frac{\Lambda_0}{3}\bar{r}^2,
\end{eqnarray}
where $\beta^2_q\equiv{q^2G}/{4\pi\epsilon_0}$
and $q$ is the overall charge of the black hole thus formed. By defining $A\equiv 4\pi\gamma^2$ as the area of the matter distribution we obtain
\begin{eqnarray}
\label{charge}
\lambda=\frac{q^2}{2A^2\epsilon_0}+\frac{E_\gamma}{3}.
\end{eqnarray}
This latter result constrains the interaction parameter with $\lambda>E_\gamma/3$. 
Furthermore, considering its low energy
domain, in the following we will set $\Lambda_0\rightarrow 0$ in the collapse scale. In this case, it is easy to see that $E_d>2(3\lambda-E_\gamma)$ is a sufficient condition for bounded configurations so that the matter distribution keeps bouncing between $\bar{r} = \gamma$ and $\bar{r} = \gamma a_{min}$.
In Fig. 1, we show a numerical simulation which illustrates a simple domain for bounded configurations.
\begin{figure}
\includegraphics[width=7cm,height=5cm]{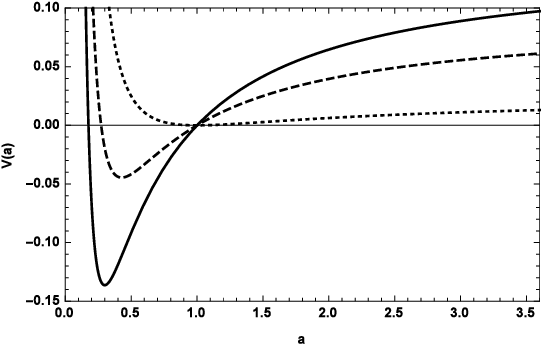}
\caption{The potential $V(a)$ for several values of $E_d$. In the above we have fixed $\lambda=0.05$,
$E_\gamma=10^{-4}$, $\kappa=1$ and $\Lambda_0=0$.
Solid and dashed lines correspond to
$E_d=1.0$ and $0.7$, respectively. In both cases, nonsingular bounded configurations are allowed. The dotted line is connected 
to 
$E_d=2(3\lambda-E_\gamma)$ -- the lower limit for bounded configurations -- so that nonsingular bounded configurations are absent.}
\label{fig1}
\end{figure}
Moreover, as long as $G^2 M^2>\beta_q^2$, it can 
be shown that the exterior spacetime has an exterior event horizon $\bar{r}_{+}$ and an interior
Cauchy horizon $\bar{r}_{-}$ given by 
\begin{eqnarray}
\label{horizons}
\bar{r}_{+}=GM+\sqrt{G^2M^2-\beta_q^2},\\
\bar{r}_{-}=GM-\sqrt{G^2M^2-\beta_q^2}.
\end{eqnarray}
From now on we shall restrict ourselves to such bounded configurations in which $\gamma a_{min}<\bar{r}_{-}$.
In Fig. 2 we show the Penrose diagram for the exterior spacetime.
\begin{figure}
\includegraphics[width=0.6\textwidth, angle =270 ]{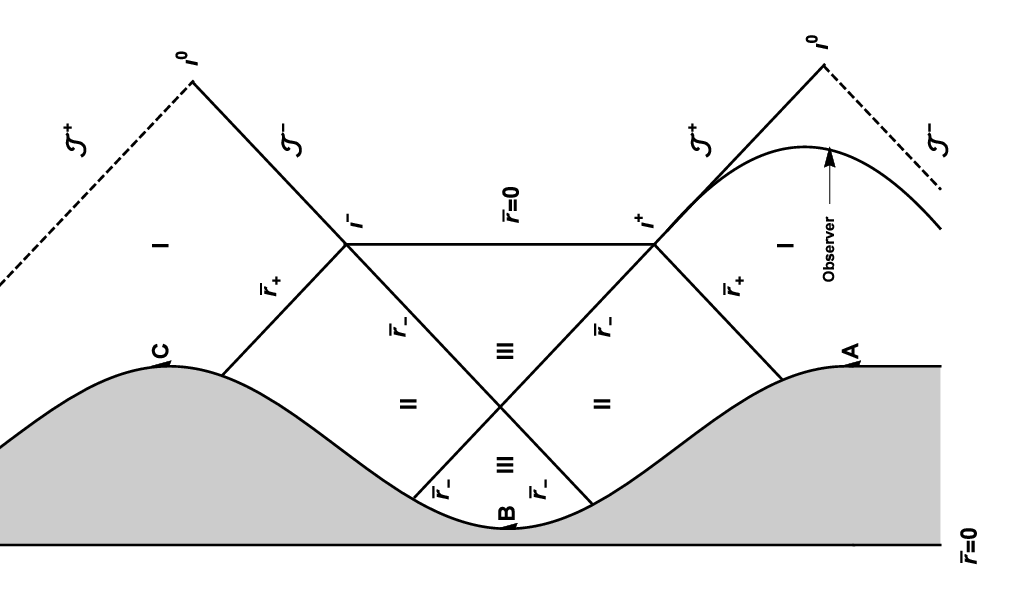}
\caption{Penrose diagram for the exterior spacetime assuming $G^2M^2 > \beta^2$. The infinite chain of asymptotically
flat regions $I$ $(\infty > \bar{r} > \bar{r}_{+})$ are connected to regions $III$ $(\bar{r}_{-} > \bar{r} > \gamma a_{min})$ by regions $II$ $(\bar{r}_{+} > \bar{r} > \bar{r}_{-})$. The shaded portion, corresponds to the
interior of the matter distribution. The line $ABC$ illustrates how the matter surface
evolves in time. Once crossed $\bar{r}_{-}$ such a surface keeps bouncing between $\bar{r} = \gamma$ and $\bar{r} = \gamma a_{min}$.}
\label{fig1}
\end{figure} 

To proceed, we now examine some important results derived from our model in the context of black hole thermodynamics.
Restricting ourselves to the natural units $G = c = \hslash = K_B = 1$, let us consider our solution in
the case of quasi-extremal black holes. Namely, let $\bar{r}_{\pm}=M_e\pm\epsilon$,
where $M_e$ is the mass of the extremal case and $\epsilon\equiv (M^2-\beta^2_q)^{1/2}\ll 1$. 
In this regime, Hawking temperature\cite{hawking} reads
\begin{eqnarray}
\label{hr2}
T_H\simeq \frac{\epsilon}{2\pi M_e^2}.
\end{eqnarray}
Defining the outer horizon area as $A_{outer}:=4\pi \bar{r}_{+}^2\simeq 4\pi M_e(1+2\epsilon)$,
we obtain 
\begin{eqnarray}
\label{fl}
dA_{outer}\simeq \frac{4}{T_H}\Big(dM- d\beta_q\Big).
\end{eqnarray}
We can therefore associate the horizon area of the
quasi-extremal black hole with the geometrical entropy $S_{\rm geom}$ 
through the relation
\begin{eqnarray}
\label{sgeom}
dS_{\rm geom} \simeq \frac{1}{4} dA_{outer} \simeq \frac{1}{T_H}\Big(dM- d\beta_q\Big).
\end{eqnarray}
Equation (\ref{sgeom}) corresponds to an
extended first law of Bekenstein\cite{beken} with an extra work term connected to charge variation.

We are now led to construct a statistical model for the
collapsed matter  
which
may be connected to the geometrical thermodynamics discussed above. 
To this end let us consider the first integral (\ref{eqn2}) with the initial conditions (\ref{ic}). Once the
surface of matter distribution is defined by $\bar{r} = \gamma a$, we can define the momentum (per unit of mass)
as $p_{\bar{r}} := \gamma \dot{a}$. In this case, 
the expansion of (\ref{eqn2}) in a neighbourhood of the extremal configuration 
furnish the Hamiltonian constraint
%
%
\begin{eqnarray}
\label{exp}
{\cal H}=\frac{p^2_R}{2}+\frac{1}{2}\Big(\frac{1}{\epsilon^2+M_e^2}\Big)R^2+ \frac{k\gamma^2-1}{2}\simeq 0,
\end{eqnarray}
where we redefine de radial coordinate by $R:=\bar{r}-M_e$. The first integral above shows that 
the interior particles of the matter distribution must oscillate with a frequency $\omega$
given by
\begin{eqnarray}
\label{fr}
\omega=\frac{1}{\sqrt{\epsilon^2+M_e^2}}\simeq \frac{1}{M_e^2}\Big(1-\frac{\epsilon^2}{M_e}\Big).
\end{eqnarray}
Defining $N$ as the number of Planck masses contained in the matter distribution of the
extremal case, $N = M_e /m_{Pl}$, the approximated motion of our system can then
be interpreted as the motion of $N$ oscillators with frequency
$\omega$. The fluctuations about the extremal configuration  
will engender quantum thermal
fluctuations that will have a contribution in the following canonical partition function
\begin{eqnarray}
\label{par}
Z=\Big\{\sum_{n=0}^\infty \exp{\Big[-\Big(n+\frac{1}{2}\Big)\frac{\omega}{T_H}\Big]}\Big\}^N=\Big[\frac{\exp(\omega/ 2  T_H)}{\exp(\omega/ T_H-1)}\Big]^N\simeq \exp{\Big(-\frac{N\omega}{ 2 T_H}}\Big).
\end{eqnarray}
Therefore, the free energy reads
\begin{eqnarray}
\label{fe}
F=-{\cal R} T_H \ln{Z}\simeq \frac{N{\cal R}\omega }{2},
\end{eqnarray}
where ${\cal R}$ is an appropriate constant. The entropy of the system can
be calculated as $S = -\partial F/\partial T_H$,
resulting in
\begin{eqnarray}
\label{ent}
S_{stat}\simeq\frac{2\pi N {\cal R} }{M_e}\epsilon \rightarrow dS_{stat}\equiv \frac{1}{T_H}\Big(dM-\frac{1}{G}d\beta_q\Big)
\end{eqnarray}
where we have fixed ${\cal R} \equiv K_B N$. Here we see that $N$ arises
naturally as the analog of an Avogadro number for the internal matter distribution
of the extremal case. 
By comparing Eqs. (\ref{sgeom}) and (\ref{ent}) we are led to identify $S_{geom}$ with $S_{stat}$
once both equations reflect the first law of black hole thermodynamics.
We should remark however that the statistical entropy derived in (\ref{ent})
differs from Bekenstein's entropy $S_{geom}$ (\ref{sgeom}) by a zero temperature additive constant
which corresponds to the area of the extremal black hole. In fact, restoring the constants $K_B$, $G$ and $\hslash$, it easy to see that as long as
\begin{eqnarray}
\label{rel}
S_{geom}=\frac{K_B}{4G\hslash} A_{ext}+S_{stat},
\end{eqnarray}
one may assign a statistical model for the geometrical entropy of Bekenstein.

Notwithstanding these striking similarities, a fundamental issue can now be posed.
In accordance with usual black hole thermodynamics, the entropy of a black hole is an external 
variable connected to the
event horizon boundary. In this sense, the entropy
of the collapsed matter would be irrelevant for physical processes outside the
black hole. To conclude this essay, we devote these final lines to provide an answer to this
issue.

Let us consider the frequency $\Delta \omega$ of a particle of the oscillating
matter distribution with mass $m_{pl}$, given by (cf. (\ref{fr}))
\begin{eqnarray}
\label{dw}
\Delta\omega\simeq \frac{\epsilon^2}{G^3 M_e^3}.
\end{eqnarray}
Its associated momentum, on the other hand, reads $\Delta P = \sqrt{2 m_{pl} \hslash \Delta\omega}$. According to Heisenberg
uncertainty principle, the fluctuations in its localization satisfies $\Delta P \Delta R \geq \hslash/2$ and, using (\ref{dw}),
we obtain
\begin{eqnarray}
\label{dr}
\Delta R \geq \frac{G M_e}{\epsilon}\sqrt{\frac{G M_e \hslash}{8m_{pl}}}.
\end{eqnarray}
For illustration, by taking $M_e$ as the Chandrasekhar mass $(M_e = 1.4M_{\odot})$ we find 
that if 
$\epsilon \lesssim 10^{5}l_p$ (where $l_p$ is the Planck
length), we can ensure that the scale of the fluctuations $\Delta R \thicksim \epsilon$ so that the
quantum thermal fluctuations that give rise to the entropy (\ref{ent}) might be connected to the fluctuations of the event horizon in the quasi-extremal
case. In this sense, (\ref{rel}) validates the identification of both
entropies with the proviso that the temperature responsible for the thermal fluctuations
is identified with the Hawking temperature.

\end{document}